\documentclass[graybox]{svmult}
\usepackage{amsmath}
\usepackage{amssymb}
\usepackage{bm}
\usepackage{bbm}
\usepackage{mathptmx}       
\usepackage{helvet}         
\usepackage{courier}        
\usepackage{type1cm}        
\usepackage{xcolor}
\usepackage{makeidx}         
\usepackage{graphicx}        
\usepackage{multicol}        
\usepackage[bottom]{footmisc}
\usepackage{cite}

\makeindex             

\begin{document}

\title*{Supracentrality Analysis of Temporal Networks with Directed Interlayer Coupling}
\titlerunning{Supracentrality with Directed Coupling}
\author{Dane Taylor, Mason A. Porter, and Peter J. Mucha}
\institute{Dane Taylor \at University at Buffalo, State University of New York, \email{danet@buffalo.edu}
\and Mason A. Porter \at University of California, Los Angeles, \email{mason@math.ucla.edu}
\and Peter J. Mucha \at University of North Carolina, Chapel Hill, \email{mucha@unc.edu}
}
\maketitle

\abstract{
We describe centralities in temporal networks using a supracentrality framework to study centrality trajectories, which characterize how the importances of nodes change in time. We study supracentrality generalizations of eigenvector-based centralities, a family of centrality measures for time-independent networks that includes PageRank, hub and authority scores, and eigenvector centrality. We start with a sequence of adjacency matrices, each of which represents a time layer of a network at a different point or interval of time. Coupling centrality matrices across time layers with weighted interlayer edges yields a \emph{supracentrality matrix} $\mathbb{C}(\omega)$, where $\omega$ controls the extent to which centrality trajectories change over time. We can flexibly tune the weight and topology of the interlayer coupling to cater to different scientific applications. The entries of the dominant eigenvector of $\mathbb{C}(\omega)$ represent \emph{joint centralities}, which simultaneously quantify the importance of every node in every time layer.  Inspired by probability theory, we also compute \emph{marginal} and \emph{conditional centralities}. We illustrate how to adjust the coupling between time layers to tune the extent to which nodes' centrality trajectories are influenced by the oldest and newest time layers. We support our findings by analysis in the limits of small and large $\omega$.
}


\section{Introduction}\label{sec:intro}

Quantifying the importances of nodes through the calculation of `centrality' measures is a central topic in the study of networks \cite{newman2010}. It is important in numerous and diverse applications, including identification of influential people \cite{bonacich1972,faust1997centrality,borgatti1998network,kempe2003}, ranking web pages in searches \cite{Brin1998conf,pagerank,kleinberg1999}, ranking teams and individual athletes in sports \cite{monthly,saavedra2010,chartier2013}, identification of critical infrastructures that are susceptible to congestion or failure \cite{holme2003congestion,guimera2005worldwide}, quantifying impactful judicial documents \cite{leicht-citation2007,fowler2007b,fowler2008} and scientific publications \cite{bergstrom2008eigenfactor}, revealing drug targets in biological systems \cite{jeong2001lethality}, and much more.

Because most networks change in time \cite{holme2012temporal,holme2013,holme2015}, there is much interest in extending centralities to temporal networks \cite{liao2017ranking}. Past efforts have generalized quantities such as betweenness centrality \cite{tang2010,kim2012b,williams2015,alsayed2015,fenu2015}, closeness centrality \cite{tang2010,pan2011,kim2012b,williams2015}, Bonacich and Katz centrality \cite{lerman2010centrality,Grindrod_Higham_2014}, win/lose centrality \cite{motegi2012}, communicability \cite{grindrod2011communicability,estrada2013,Grindrod_Higham_2013,fenu2015,chen2016dynamic,arrigo2017sparse}, dynamic sensitivity \cite{huang2017dynamic}, coverage centrality \cite{taro2015}, PageRank \cite{walker2007ranking,Mariani2016,you2015distributed,rossi2012,mariani2015}, and eigenvector centrality \cite{praprotnik2015spectral,huang2017centrality,flores2018eigenvector}.
{A common feature of these extensions is that they illustrate the importance of using methods that are designed explicitly for temporal networks, as opposed to various alternatives: aggregating a temporal network into a single `time-independent' network; independently analyzing the temporal network at different instances in time; or binning the temporal network into time windows and analyzing those windows independently. In the first case, it is not even possible to study centrality trajectories (i.e., how centrality changes over time).}

Because one can derive many centralities by studying walks
on a network, some of the above temporal generalizations of centrality involve the analysis of so-called `time-respecting paths'
 \cite{kossinets2008structure,kostakos2009temporal}. 
 There are multiple ways to define a time-respecting path, including the possibility of allowing multiple edge traversals per time step for a discrete-time temporal network.
There are also multiple ways to quantify the length of a time-respecting path \cite{williams2015}, 
because such a length can describe the number of edges that are traversed by a path, latency between the initial and terminal times of a path, or a combination of these ideas. In particular, it is necessary to make choices even to define a notion of a `shortest path' (from which one can formulate several types of centrality). 
Consequently, some of the diversity in the various temporal generalizations of centrality measures arises from the diversity in defining and measuring the length of a time-respecting path.

\begin{figure}[t]
\centering{
\includegraphics[width=.9\linewidth]{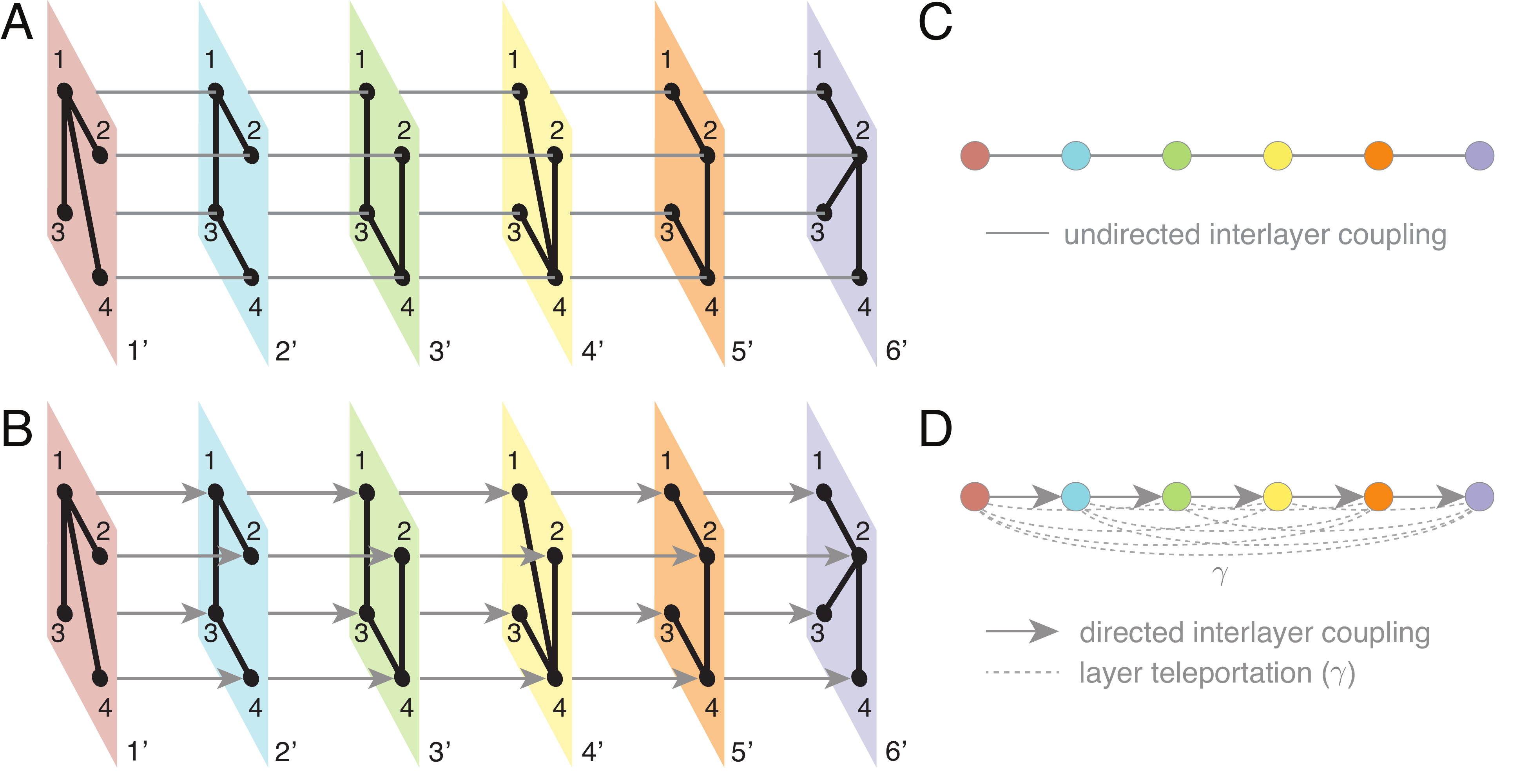}}
\vspace{-.2cm}
\caption{{\bf Multiplex-network representations of a discrete-time temporal network.} Given a temporal network with $N=4$ nodes and $T=6$ times, we represent the network at each time by a `time layer' with adjacency matrix  ${\bf A}^{(t)} \in\mathbb{R}^{N\times N}$ for $t\in\{1,\dots,T\}$. 
(A,B) 
We represent the network
as a multiplex network by coupling the layers with `interlayer edges' (gray edges) that we encode in an interlayer-adjacency matrix $\tilde{\bm{A}}  \in\mathbb{R}^{T\times T}$. Panel (A) illustrates interlayer coupling in the form of an undirected chain, and panel (B) depicts directed coupling between layers. In panels (C) and (D), we show visualizations of the networks that are associated with $\tilde{\bm{A}}$ for panels (A) and (B), respectively. In panel (D), there are directed interlayer edges between consecutive time layers, so these interlayer edges respect the direction of time.
Additionally, we  construct connections of weight $\gamma>0$ between corresponding nodes from all pairs of layers to ensure that $\tilde{\bm{A}}$ corresponds to a strongly connected network, which in turn ensures that the centralities are positive and unique. 
By analogy to `node teleportation' in PageRank \cite{gleich2014}, we refer to this coupling as `layer teleportation'.
}
\label{fig:toy1}
\end{figure}

In the present work, we examine a notion \emph{supracentrality} \cite{taylor2017eigenvector,taylor2019tunable}, which one can calculate by representing a temporal network as a sequence of network layers and coupling those layers to form a multilayer network (specifically, a multiplex network \cite{kivela2014,whatis2018}). See Fig.~\ref{fig:toy1} for illustrative examples. The sequence of network layers, which constitute \emph{time layers}, can represent a discrete-time temporal network at different time instances or a continuous-time network in which one bins (i.e., aggregates \cite{taylor2017super}) the network's edges to form a sequence of time windows with interactions in each window. 
This approach is motivated by the use of a multiplex-network representation to detect 
communities in temporal networks through maximization of multilayer modularity \cite{muchaporter2010,bassett2013,weir2017post,roxana2019}. 
We note in passing that there is also widespread interest in generalizing centrality measures to multilayer networks more generally \cite{magnani2011ml,de2013b,battiston2014structural,tavassoli2016most,magnani2013combinatorial,sole2014centrality,chakraborty2016cross,sole2016random,spatocco2018new,rahmede2017centralities,tudisco2018node,sola2013eigenvector,deford2017new,deford2017multiplex,ng2011multirank,Halu_Mondragon_Panzarasa_Bianconi_2013,ding2018centrality}. 

{Our  supracentrality framework generalizes a family of centralities for time-independent networks called \emph{eigenvector-based} centralities, which are defined by the property of calculating centralities as the entries of an eigenvector (the so-called `dominant' eigenvector) that corresponds to the largest-magnitude eigenvalue (the `dominant' eigenvalue) of a \emph{centrality matrix} ${C}(\mathbf{A})$, which one defines by some function of a network adjacency matrix $\mathbf{A}$. Different choices for the centrality matrix recover some of the most popular centrality measures, including eigenvector centrality (by using ${C}(\mathbf{A}) = \mathbf{A}$) \cite{bonacich1972}, hub and authority scores (by using ${C}(\mathbf{A}) = \mathbf{A}\mathbf{A}^T$ for hubs and $\mathbf{A}^T\mathbf{A}$ for authorities) \cite{kleinberg1999}, and PageRank \cite{pagerank} (see Sec.~\ref{sec:eig_central}).}
Given a discrete-time temporal network in the form of a sequence of adjacency matrices ${\bf A}^{(t)} \in\mathbb{R}^{N\times N}$ for $t\in\{1,\dots,T\}$, where $A_{ij}^{(t)}$ denotes a directed edge from entity $i$ to entity $j$ in time layer $t$, examining supracentralities involves two steps: 
\begin{enumerate}
\item  Construct a {supracentrality matrix} $\mathbb{C}(\omega)$, which couples centrality matrices ${C}({\bf A}^{(t)})$ of the individual time layers $t=1$, $t = 2$, $t = 3$, \ldots
\item  Compute and interpret the dominant eigenvector of $\mathbb{C}(\omega)$. 
\end{enumerate}
For a temporal network with $N$ nodes and $T$ time layers, $\mathbb{C}(\omega)$ is a square matrix of size $NT\times NT$. We require the set of nodes to be the same for all time layers. 
However, it is easy to accommodate the appearance and disappearance of nodes by including extra instances of the entities in layers in which they otherwise do not appear (but without including any associated intralayer edges).
The parameter $\omega$ scales the weights of the interlayer coupling to control the strength of the connection between time layers. It thus provides a `tuning knob' to control how rapidly centrality trajectories can change over time. 


An important aspect of the first step is that one must choose a topology to couple layers to each other. To do this, we define an interlayer-adjacency matrix $\tilde{\bm{A}}  \in\mathbb{R}^{T\times T}$, where the entry $\tilde{{A}}_{tt'}$ encodes the coupling from time layer $t$ to time layer $t'$. In Fig.~\ref{fig:toy1}, we illustrate two possible choices for coupling the time layers. In the upper row, $\tilde{\bm{A}}  \in\mathbb{R}^{T\times T}$ encodes an undirected chain, which couples the time layers with \emph{adjacent-in-time} coupling but neglects the directionality of time. In the lower row, by contrast, we couple the time layers with a directed chain that reflects the directionality of time. In addition to the directed, time-respecting edges, Fig.~\ref{fig:toy1}(D) also illustrates that we include weighted, undirected edges between corresponding nodes in all pairs of layers.
This implements `layer teleportation', which is akin to the well-known `node teleportation' of the PageRank algorithm \cite{gleich2014}. Similar to the motivation for node teleportation, layer teleportation ensures that supracentralities are well-behaved (specifically, that they are positive and unique).

The second step of our supracentrality framework involves studying the dominant right eigenvector of the supracentrality matrix $\mathbb{C}(\omega)$, which characterizes the \emph{joint centrality} of each node-layer pair $(i,t)$---that is, the centrality of node $i$ in time layer $t$---and thus reflects the importances of both node $i$ and layer $t$. From the joint centralities, one can calculate
\emph{marginal centralities} for only the nodes (or only the time layers). One can also calculate \emph{conditional centralities} that measure a node's centrality at time $t$ relative only to the other nodes' centralities in that particular time layer $t$. These 
concepts, which are inspired by ideas from probability theory, allow one to develop a rich characterization for how node centralities change over time. 

In this chapter, we describe the supracentrality framework that we developed in \cite{taylor2017eigenvector,taylor2019tunable} and extend these papers with further numerical explorations of how interlayer coupling topology affects supracentralities. We apply this approach to a data set, which we studied in \cite{taylor2017eigenvector} and is available at \cite{MGP_data}, that encodes the graduation and hiring of Ph.D. recipients between mathematical-sciences 
doctoral programs in the United States. We focus our attention on five top universities and examine how they are affected by the value of $\omega$ and the choice of $\tilde{\bm{A}}$. Specifically, we compare the two strategies for interlayer coupling in Fig.~\ref{fig:toy1}, and we explore the effect of reversing the directions of all directed edges. Our experiments reveal how to use $\omega$ and $\tilde{\bm{A}}$ to tune the extent to which centrality trajectories of nodes are influenced by the oldest time layers, the newest time layers, and the direction of time.

\vspace{-.3cm}

\section{Background Information}\label{sec:back}

We now give some background information on multiplex networks and eigenvector-based centralities. Our supracentrality framework involves representing a temporal network as a multiplex network (see Sec.~\ref{sec:mux}).
In Sec.~\ref{sec:eig_central}, we review eigenvector-based centrality measures.



\subsection{Analysis of Temporal Networks with Multiplex-Network Representations}\label{sec:mux}

We study discrete-time temporal networks, for which we provide a formal definition. 

\begin{definition}[Discrete-Time Temporal Network]\label{def:time}
A \emph{discrete-time temporal network} consists of a set $\mathcal{V}=\{1,\dots,N\}$ of nodes and sets $\mathcal{E}^{(t)}$ of weighted edges that we index (using $t$) in a sequence of network layers. We denote such a network either as $\mathcal{G}(\mathcal{V},\{\mathcal{E}^{(t)}\})$ or by the sequence $\{\mathbf{A}^{(t)}\}$ of adjacency matrices, where ${A}_{ij}^{(t)}=w^t_{ij}$ if $(i,j,w^t_{ij})\in\mathcal{E}^{(t)}$ and $ {A}_{ij}^{(t)}=0$ otherwise. 
\end{definition}


As we illustrated in Fig.~\ref{fig:toy1}, we represent a discrete-time temporal network as a multiplex network with weighted and possibly directed coupling between the time layers. 
We restrict our attention to the following type of multiplex network.


\begin{definition}[Uniformly and Diagonally-Coupled (i.e., Layer-Coupled) Multiplex Network]\label{def:mux}
Let $\mathcal{G}(\mathcal{V},\{\mathcal{E}^{(t)}\},\tilde{\mathcal{E}})$ be a $T$-layer multilayer network with node set $\mathcal{V}=\{1,\dots,N\}$ 
and 
interactions between node-layer pairs that are encoded by the sets $\{\mathcal{E}^{(t)}\}$ of weighted edges, where $(i,j,w^t_{ij})\in\mathcal{E}^{(t)}$ if and only if there is an edge $(i,j)$ with weight $w^t_{ij}$ in layer $t$. The set $\tilde{\mathcal{E}}=\{(s,t,\tilde{w}_{st})\}$ encodes the topology and weights for coupling separate 
instantiations
of the same node between a pair, $(s,t)\in\{1,\dots, T\}\times \{1,\dots, T\}$, of layers. Equivalently, one can encode a multiplex network as a set $\{\mathbf{A}^{(t)}\}$ of adjacency matrices, such that $ {A}_{ij}^{(t)}=w^t_{ij}$ if $(i,j,w^t_{ij})\in\mathcal{E}^{(t)}$ and $ {A}_{ij}^{(t)}=0$ otherwise, along with an interlayer-adjacency matrix $\tilde{\bm{A}}$ with entries $\tilde{A}_{st} = \tilde{w}_{st}$ if $(s,t,\tilde{w}_{st}) \in \tilde{\mathcal{E}}$ and $\tilde{A}_{st}^{(t)}=0$ otherwise.
\end{definition}

The  coupling in Definition \ref{def:mux} is `diagonal' in that the only interlayer edges are ones that couple a node in one layer with that same node in another layer. It is `uniform' in that the coupling between two layers is identical for all nodes in those two layers. A multilayer network with both  conditions is called `layer-coupled' \cite{kivela2014}. 
%

As we illustrate in Fig.~\ref{fig:toy1}, we focus our attention on two choices for coupling time layers: 

(A) $\tilde{\bm{A}}$ encodes an undirected chain:
\begin{equation}\label{eq:undir}
	\tilde{A}_{tt'} = 
\left\{\begin{array}{rl}
	1\,,& \, |t'-t|= 1\,, \\
0 \,,& \, \text{otherwise} \,; \\
\end{array} \right. 
\end{equation}

(B) $\tilde{\bm{A}}$ encodes a directed chain with layer teleportation:
\begin{equation}\label{eq:tele}
	\tilde{A}_{tt'} = 
\left\{\begin{array}{rl}
	1+\gamma\,,& \, t'-t= 1\,, \\
\gamma \,,& \, \text{otherwise} \,, \\
\end{array} \right. 
\end{equation}
where $\gamma>0$ is the \emph{layer-teleportation probability}. In Sec.~\ref{sec:data}, we compare the effects on centrality trajectories of these two choices for $\tilde{\bm{A}}$.


\subsection{Eigenvector-Based Centrality for Time-Independent Networks}\label{sec:eig_central}

Arguably the most notable---and certainly the most profitable---type of centrality is PageRank, which provided the mathematical foundation for the birth of the web-search algorithm of the technology giant Google \cite{Brin1998conf,pagerank,gleich2014}. PageRank quantifies the importances of nodes in a network (e.g., a directed network that encodes hyperlinks between web pages) by computing the dominant eigenvector of 
the `PageRank matrix' (or `Google matrix' \cite{langville2006})
\begin{align} \label{eq:google}
	\mathbf{C}^{(PR)} &= {\sigma} \mathbf{A}^T\mathbf{D}^{-1} + (1-{\sigma})N^{-1}\mathbf{1}\mathbf{1}^T\,,
\end{align}
where $N$ is the number of nodes, $\mathbf{1}=[1,\dots,1]^T$ is a length-$N$ vector of ones, and $\mathbf{A}$ is an adjacency matrix in which each entry $A_{ij}$ encodes a directed (and possibly weighted) edge from node $i$ to node $j$. The matrix $\mathbf{D}=\text{diag}[d_1^{\mathrm{out}},\dots,d_N^{\mathrm{out}}]$ is a diagonal matrix that encodes the node out-degrees $d_i^{\mathrm{out}} = \sum_j A_{ij}$. 

The PageRank matrix's dominant right eigenvector is a natural choice for ranking nodes, as it encodes a random walk's stationary distribution (which estimates the fraction of web surfers on each web page in the context of a web-search engine\footnote{
Note that PageRank has had intellectual impact well beyond web searches \cite{gleich2014}.}).  
The term $\mathbf{A}^T\mathbf{D}^{-1}$ is a transition matrix that operates on column vectors that encode the densities of random walkers \cite{masuda2017}.  The term $N^{-1}\mathbf{1}\mathbf{1}^T$ is a \emph{teleportation matrix}; it represents a transition matrix in a network with uniform all-to-all coupling between nodes.
The \emph{teleportation parameter} $\sigma\in(0,1)$ implements a linear superposition of the two transition matrices and yields an irreducible matrix, even when
the transition matrix $\mathbf{A}^T\mathbf{D}^{-1}$ is reducible. 
Because we introduced the concept of layer teleportation  in Sec.~\ref{sec:mux}, we henceforth refer to the traditional teleportation in PageRank as `node teleportation'.

It is common to define the PageRank matrix as the transpose of Eq.~\eqref{eq:google}; in that case, one computes the dominant left  eigenvector instead of the dominant right one. However, we use the right-eigenvector convention to be consistent with a broader class of centrality measures called {`eigenvector-based centralities'},
%
in which one encodes node importances in the elements of the dominant eigenvector of some {centrality matrix}.
In addition to PageRank, prominent examples of eigenvector-based centralities include (vanilla) eigenvector centrality \cite{bonacich1972} 
and hub and authority (i.e., HITS) centralities \cite{kleinberg1999}.
We  now provide formal definitions.

{
 %
%
\begin{definition}[Eigenvector-Based Centrality] \label{def:EigenBased}
Let $\mathbf{C}=C(\mathbf{A})$ be a centrality matrix, which we obtain from some function $C:\mathbb{R}^{N\times N}\mapsto \mathbb{R}^{N\times N}$ of the adjacency matrix $\mathbf{A}$, of a network $\mathcal{G}(\mathcal{V},\mathcal{E})$. Consider the dominant right eigenvector $\mathbf{u}$, which satisfies 
\begin{equation}\label{eq:eigen_1}
	\mathbf{C} \mathbf{u} = \lambda_{\rm{max}} \mathbf{u} \,,
\end{equation}
where $\lambda_{\rm{max}} \in\mathbb{R}_{+}$ is the largest eigenvalue of $\mathbf{C}$. (Note that this eigenvalue is guaranteed to be positive.)
The $i$th entry $u_i$ specifies the \emph{eigenvector-based centrality} of node $i\in\mathcal{V}$ that is associated with the function $C$.
\end{definition}


\begin{definition}[PageRank \cite{pagerank,gleich2014}]\label{def:PageRank}
When $\mathbf{C}$ is given by Eq.~\eqref{eq:google}, we say that Eq.~\eqref{eq:eigen_1} yields \emph{PageRank} centralities $\{u_i^{(\mathrm{PR})}\}$.  
\end{definition}

\begin{remark}
It is also common to compute PageRank centralities from a left eigenvector \cite{gleich2014}. In the present paper, we use a right-eigenvector formulation to be consistent with the other eigenvector-based centralities. One can recover the other formulation by taking the transpose of Eq.~\eqref{eq:eigen_1}.
\end{remark}




}


\section{Supracentrality Framework}\label{sec:method}
%

We now describe the supracentrality framework that we presented in \cite{taylor2019tunable}. The present formulation generalizes our formulation of supracentrality from \cite{taylor2017eigenvector} that  required  interlayer coupling to take the form of an undirected chain. (See the top row of Fig.~\ref{fig:toy1}.) To aid our presentation, we summarize our mathematical notation in Table~\ref{table:notation}.

\begin{table}[h]
\caption{Summary of our mathematical notation for objects with different dimensions.}
 {\centering
\begin{tabular}{c c c} 
 \hline 
Typeface & Class & Dimension  \\ [0.5ex]
\svhline 
$\mathbb{M}$ & matrix & $NT\times NT$   \\ 
$\mathbf{M}$ & matrix & $N\times N$  \\
$\bm{M}$ & matrix & $T\times T$   \\
$\mathbbm{v}$ & vector & $NT\times 1$  \\
$\mathbf{v}$ & vector & $N\times 1$   \\ 
$\bm{v}$ & vector & $T\times 1$   \\ 
$M_{ij}$ & scalar & 1   \\ 
$v_i$ & scalar & 1   \\
[1ex] 
\hline 
\end{tabular}\\
\label{table:notation}} 
\end{table}

\subsection{Supracentrality Matrices}\label{sec:supra}
%
We first describe a supracentrality matrix from \cite{taylor2019tunable}.

\begin{definition}[Supracentrality Matrix]\label{def:Supracentrality}
Let $\{{\bf C}^{(t)}\}$ be a set of $T$ centrality matrices for a discrete-time temporal network with a common set $\mathcal{V}=\{1,\dots,N\}$ of nodes; and assume that ${C}^{(t)}_{ij} \ge 0 $. 
Let $\tilde{\bm{A}}$, with entries $\tilde{A}_{ij}\ge 0$, be a $T\times T$ interlayer-adjacency matrix that encodes the interlayer couplings.
We define a family of \emph{supracentrality matrices} $\mathbb{C}(\omega)$, which are parameterized by the interlayer-coupling strength $\omega \ge 0$, of the form 
\begin{align}
	{\mathbb{C}} (\omega) &=  \hat{\mathbb{C}} +  \omega \hat{\mathbb{A}}\,
	= \left[ \begin{array}{cccc} 
 \mathbf{C}^{(1)} & {\bf 0} &  {\bf 0}&\dots\\ 
{\bf 0} & \mathbf{C}^{(2)} & {\bf 0}& \dots\\ 
{\bf 0} &  {\bf 0} &  \mathbf{C}^{(3)} &\ddots\\
 \vdots&   \vdots & \ddots&\ddots\\
 \end{array}
 \right] + \omega
 \left[ \begin{array}{cccc} 
  \tilde{A}_{11} \mathbf{I}~&  \tilde{A}_{12} \mathbf{I} ~& \tilde{A}_{13} \mathbf{I}  &\dots \\ 
\tilde{A}_{21} \mathbf{I} ~& \tilde{A}_{22} \mathbf{I} ~& \tilde{A}_{23} \mathbf{I}& \dots \\ 
 \tilde{A}_{31} \mathbf{I} ~& \tilde{A}_{32}  \mathbf{I} ~& \tilde{A}_{33} \mathbf{I} ~&\dots\\
 \vdots  &   \vdots&\vdots&\ddots\\\\
 \end{array}
 \right] \, , \label{eq:supracentrality}
\end{align}
where $\hat{\mathbb{C}} = \text{\rm diag}[ \mathbf{C}^{(1)},\dots, \mathbf{C}^{(T)}]$ and $\hat{\mathbb{A}}=\tilde{\bm{A}}\otimes \bf{I}$ is the Kronecker product of $\tilde{\bm{A}}$ and $\bf{I}$.
\end{definition}

For layer $t$, the matrix ${\bf C}^{(t)}$ can represent any matrix whose dominant eigenvector is of interest. In our discussion, we focus on PageRank {(see Definition \ref{def:PageRank}), but one can alternatively choose eigenvector centrality \cite{bonacich1972}, hub and authority centralities \cite{kleinberg1999}, or something else.}


The $NT\times NT$ supracentrality matrix $\mathbb{C}(\omega) $ encodes the effects of two distinct types of connections: the layer-specific centrality entries $\{ {C}^{(t)}_{ij}\}$ in the diagonal blocks relate centralities between nodes in layer $t$; and entries in the off-diagonal blocks encode coupling between layers. The matrix $\hat{\mathbb{A}}=\tilde{\bm{A}}\otimes \bf{I}$ implements uniform and diagonal coupling. The matrix $\bf{I}$ encodes diagonal coupling; and any two layers $t$ and $t'$ are uniformly coupled, because all interlayer edges between them have the identical weight $\omega\tilde{A}_{tt'}$. 

\subsection{Joint, Marginal, and Conditional Centralities}\label{sec:joint}
%


As we indicated earlier, we study the dominant right eigenvalue equation for   supracentrality matrices. That is, we solve the eigenvalue equation
\begin{equation} \label{eq:eig_eq}
	\mathbb{{C}}(\omega)\mathbbm{v}(\omega) = \lambda_{\rm{max}}(\omega)\mathbbm{v}(\omega)\,,
\end{equation}
and we interpret entries in the dominant right eigenvector $\mathbbm{v}(\omega)$ as scores that measure the importances of node-layer pairs $\{(i,t)\}$. Because the vector $\mathbbm{v}(\omega)$ has a block form---its first $N$ entries encode the joint centralities for layer $t=1$, its next $N$ entries encode the joint centralities for layer $t=2$, and so on---it is useful to reshape $\mathbbm{v}(\omega)$ into a matrix.

\begin{definition}[Joint Centrality of a Node-Layer Pair \cite{taylor2017eigenvector}]\label{def:joint}
Let $\mathbb{C}(\omega)$ be a supracentrality matrix given by Definition~\ref{def:Supracentrality}, and let $\mathbbm{v}(\omega)$ be its dominant right eigenvector. We encode the \emph{joint centrality} of node $i$ in layer $t$ via the $N\times T$ matrix ${\bf W}(\omega)$ with entries 
\begin{equation}
	W_{it}(\omega)  = \mathbbm{v}_{N(t-1) + i}(\omega)\, . \label{eq:joint2}
\end{equation}
We refer to $W_{it}(\omega)$ as a `joint centrality' because it reflects the importance both of node $i$ and of layer $t$.
\end{definition}

\begin{definition}[Marginal Centralities of Nodes and Layers \cite{taylor2017eigenvector}]\label{def:marg}
Let ${\bf W}(\omega) $ encode the joint centralities given by Definition \ref{def:joint}. 
We define the \emph{marginal layer centrality} (MLC) and \emph{marginal node centrality} (MNC), respectively, by
\begin{align}
	 {x}_{t}(\omega) &= \sum_{i}W_{it}(\omega)\,,\nonumber\\
	 	\hat{x}_{i}(\omega) &= \sum_{t} W_{it}(\omega) \, . \label{eq:marg2}  
\end{align}
\end{definition}


\begin{definition}[Conditional Centralities of Nodes and Layers \cite{taylor2017eigenvector}]\label{def:cond}
Let $\{W_{it}(\omega)\}$ be the joint centralities given by Definition \ref{def:joint}; and let $\{{x}_{t}(\omega)  \}$ and $\{\hat{x}_{i}(\omega)  \}$, respectively, be the marginal layer and node  centralities given by Definition \ref{def:marg}.  We define the \emph{conditional centralities} of nodes and layers by
\begin{align}
	Z_{it} (\omega)&= W_{it}(\omega)/ {x}_t(\omega)\,, \nonumber\\
	\hat{Z}_{it}(\omega) &= W_{it}(\omega)/\hat{x}_i(\omega)\,,
\label{eq:cond}
\end{align}
where $Z_{it} (\omega)$ gives the centrality of node $i$ conditioned on layer $t$ and $\hat{Z}_{it} (\omega)$ gives the centrality of layer $t$ conditioned on node $i$. The quantity $Z_{it}(\omega)$ indicates the importance of node $i$ relative just to the other nodes in layer $t$.
\end{definition}


We ensure that the supracentralities are well-defined (i.e., unique, positive, and finite) with the following theorem.


\begin{theorem}[Uniqueness and Positivity of Supracentralities \cite{taylor2019tunable}] \label{thm:unique}
Let $\mathbb{C}(\omega)$ be a supracentrality matrix given by Eq.~\eqref{eq:supracentrality}. Additionally, suppose that $\tilde{\bm{A}}$ is an adjacency matrix for a strongly connected graph and that 
$\sum_t \mathbf{C}^{(t)}$ is an irreducible, nonnegative matrix. It then follows that $\mathbb{C}(\omega)$ is irreducible, nonnegative, and has a simple largest positive eigenvalue $\lambda_{\rm{max}}(\omega)$, with corresponding left eigenvector $\mathbbm{u}(\omega)$ and right eigenvector $\mathbbm{v}(\omega)$ that are each unique and positive. The centralities $\{W_{it}(\omega)\}$, $\{x_i(\omega)\}$, $\{\hat{x}_t(\omega)\}$, $ \{Z_{it}(\omega)\}$, and $ \{\hat{Z}_{it}(\omega)\}$ are then positive and finite.
If we also assume that $\mathbb{C}(\omega)$ is aperiodic, it follows that $\lambda_{\rm{max}}(\omega)$ is a unique dominant eigenvalue.
\end{theorem}



\begin{figure}[h]
\centering
\includegraphics[width=.5\linewidth]{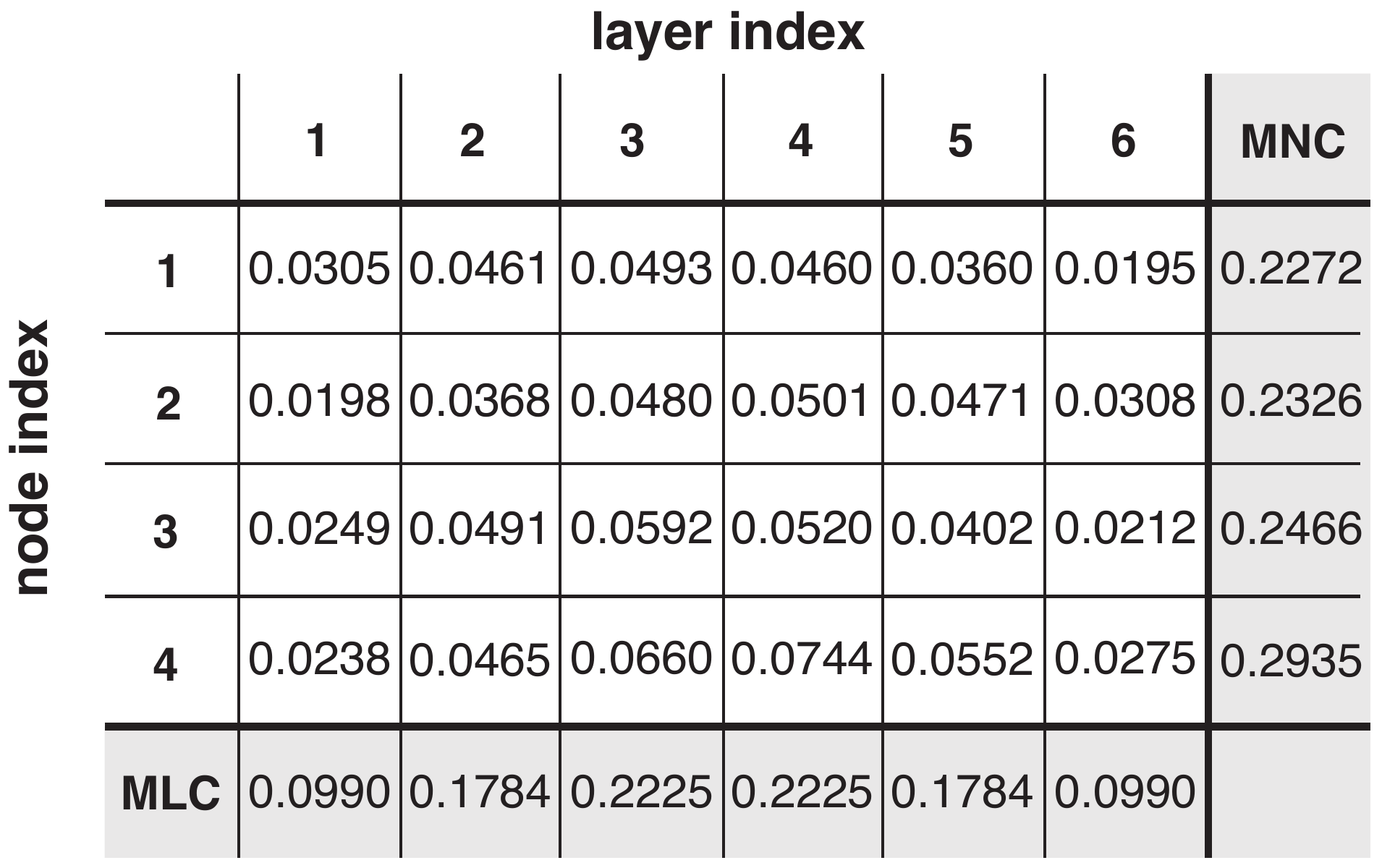}
\caption{Joint centralities $\{W_{it}(\omega)\}$ of Definition~\ref{def:joint} (white cells), with corresponding marginal layer centralities (MLC) $\{{x}_{t}(\omega)\}$ and marginal node centralities (MNC) $\{\hat{x}_{i}(\omega)\}$ from Definition~\ref{def:marg} (gray cells), for the network 
in panel (A)   of Fig.~\ref{fig:toy1} with $\omega=1$. The 
centrality matrices of the layers are PageRank centrality matrices (see Eq.~\eqref{eq:google}) with a node-teleportation parameter of $\sigma=0.85$. 
}
\label{fig:centralities_toy1}
\end{figure}

In Fig.~\ref{fig:centralities_toy1}, we show the joint and marginal centralities for the network in panel (A).
We have normalized the vector $\mathbbm{v}(\omega)$ using the 1-norm.

\section{Application to a Ph.D. Exchange Network}\label{sec:data}
%

We apply our supracentrality framework to study centrality trajectories for a temporal network that encodes the graduation and hiring of mathematicians between $N=231$ mathematical-sciences doctoral programs in the United States during the years $\{1946,\dots, 2010\}$ \cite{taylor2017eigenvector}. Each edge $A_{ij}^{(t)}$ in the temporal network encodes the number of Ph.D. recipients who graduated from university $j$ in year $t$ and subsequently supervised a Ph.D. student at university $i$.  The edge directions, where $A_{ij}^{(t)}$ is an edge from university $i$ to university $j$, point in the opposite direction 
to the flow of   people who earn their Ph.D. degrees. We define edge directions in this way to indicate that university $i$ effectively selects the output of university $j$ when they hire someone who received their Ph.D. from $j$ \cite{burris2004,myers2011,clauset2015}. With this convention for the direction of edges, $\{{\bf C}^{(t)}\}$ encodes the 
PageRank matrices of the layers; and the highest-ranking universities are 
the ones that are good sources for the flow of Ph.D. recipients. The network, which we constructed using data from the Mathematics Genealogy Project \cite{mgp}, is available at \cite{MGP_data}.

{We focus our discussion on  five U.S. universities: Harvard, Massachusetts Institute of Technology (MIT), Princeton, Stanford, and University of California at Berkeley. They  have the largest PageRank centralities  (using a node-teleportation parameter of $\sigma=0.85$) for a temporally aggregated network with adjacency matrix $\sum_t \bm{A}^{(t)}$.} 
In all of our experiments, we assume that the layers' centrality matrices are given by PageRank matrices, as defined in Eq.~\eqref{eq:google}.  As in our previous explorations \cite{taylor2017eigenvector,taylor2019tunable}, we vary the interlayer coupling strength $\omega$ to adjust how rapidly centralities change over time. In the present work, our primary focus is investigating the effects on supracentralities of undirected and directed interlayer coupling. See Eq.~\eqref{eq:undir} and Eq.~\eqref{eq:tele} for the definitions of these interlayer-coupling schemes;   see Fig.~\ref{fig:toy1} for visualizations of these two types of interlayer coupling.

\begin{figure}[t]
\includegraphics[width=1\linewidth]{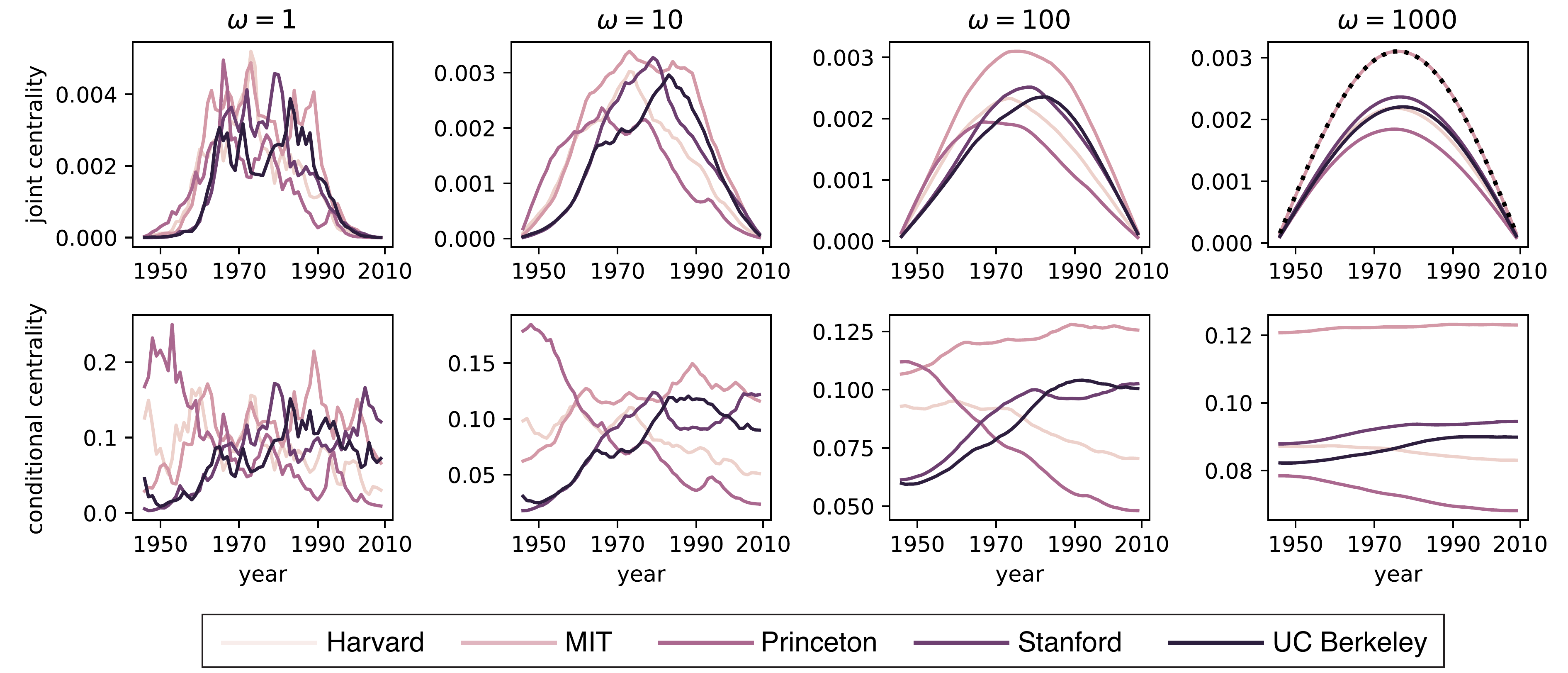}
\caption{{\bf Trajectories of node centralities using undirected interlayer coupling for mathematical-sciences Ph.D. programs at five top universities.} The top and bottom rows illustrate joint and conditional node centralities, respectively, that we compute with   centrality matrices based on PageRank with a node-teleportation parameter of $\sigma=0.85$ and undirected interlayer coupling $\tilde{\bm{A}}$ given by Eq.~\eqref{eq:undir} with $\omega\in\{1,10,100,1000\}$. The dotted black curve in the rightmost top panel is the result of an asymptotic approximation
that we present in Section \ref{sec:asym}.
}
\label{fig:undir}
\end{figure}

\begin{figure}[t]
\includegraphics[width=1\linewidth]{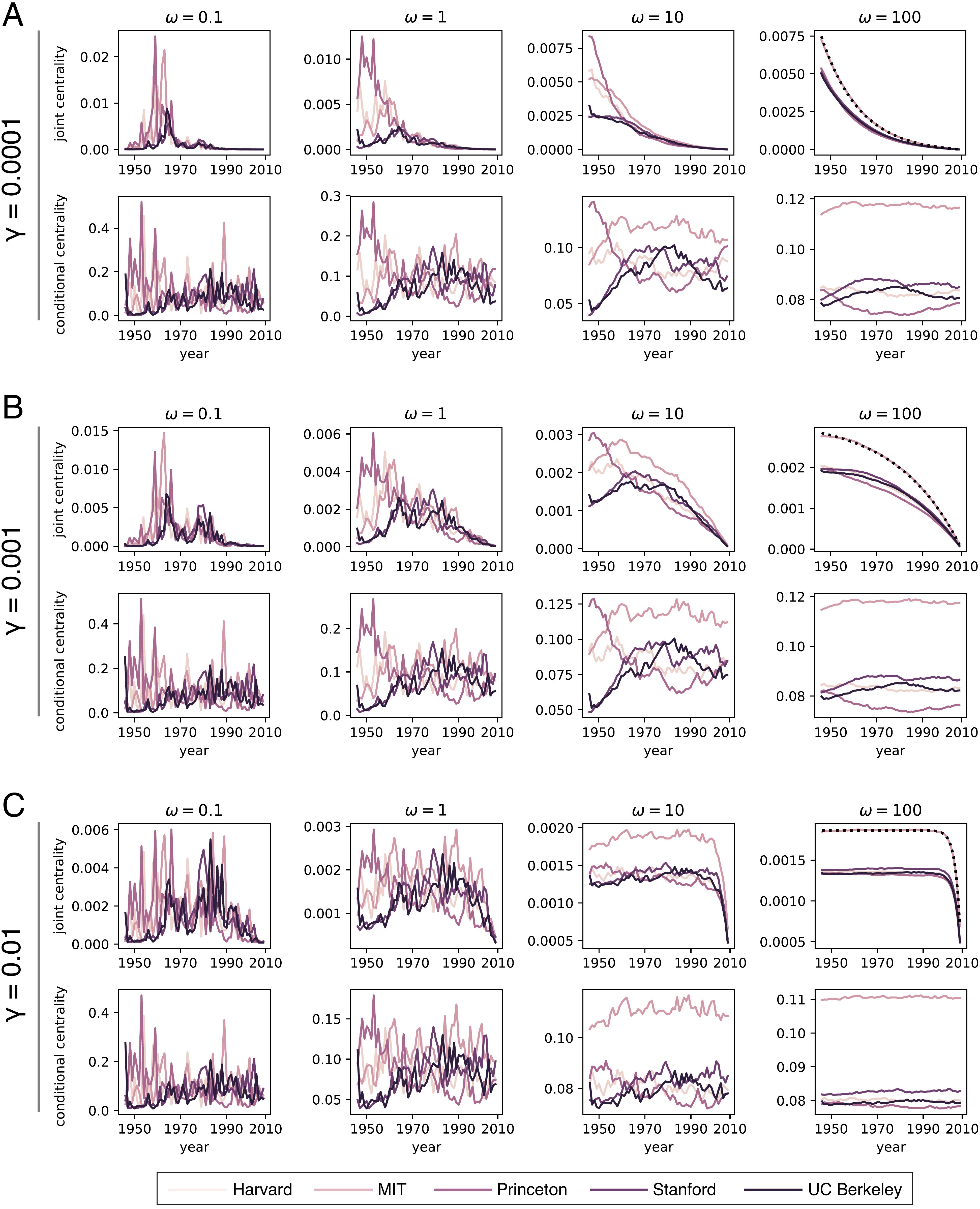}
\caption{{\bf Trajectories of node centralities using directed interlayer coupling for mathematical-sciences Ph.D. programs at five top universities.}
This figure is similar to Fig.~\ref{fig:undir}, except that the interlayer-adjacency matrix $\tilde{\bm{A}}$ is now given by Eq.~\eqref{eq:tele}, which corresponds to a directed chain with layer teleportation with rate $\gamma$. Panels (A)--(C) show results for $\gamma = 0.0001$, $\gamma = 0.001$, and $\gamma = 0.01$, respectively.  The dotted black curves in the rightmost top  subpanels of panels (A)--(C) are the result of an asymptotic approximation that we present in Section \ref{sec:asym}. For sufficiently large $\omega$ and sufficiently small $\gamma$, observe that the joint centralities 
decrease with time.
}
\label{fig:dir1}
\end{figure}


\begin{figure}[t]
\includegraphics[width=1\linewidth]{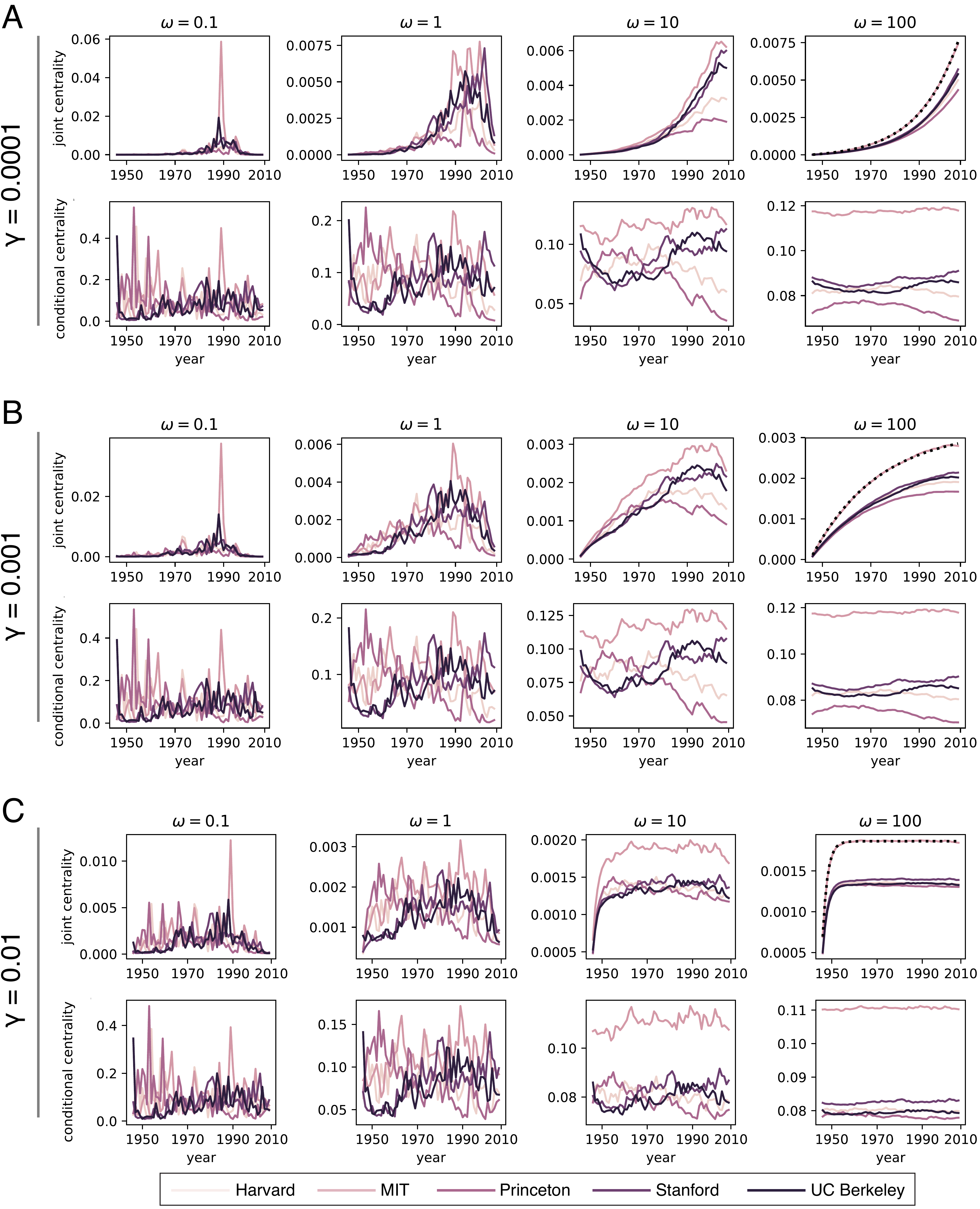}
\caption{{\bf Trajectories of node centralities using reversed directed interlayer coupling for mathematical-sciences Ph.D. programs at five top universities.}
This figure is identical to Fig.~\ref{fig:dir1}, except that  $\tilde{\bm{A}}$ is now given by the transpose of the matrix from Eq.~\eqref{eq:tele}, such that the directed chain points backwards in time. For sufficiently large $\omega$ and sufficiently small $\gamma$, observe that the joint centralities now 
increase with time.
}
\label{fig:dir2}
\end{figure}

We first consider undirected interlayer coupling, so we define $\tilde{\bm{A}}$ by Eq.~\eqref{eq:undir}. In Fig.~\ref{fig:undir}, we plot the joint and conditional centralities 
for the five universities. The columns show   results for  interlayer-coupling strengths $\omega\in\{1,10,10^2,10^3\}$. In the bottom row, we see that   progressively larger values of $\omega$ yield progressively smoother conditional-centrality trajectories. In the top row, we observe that as one increases  $\omega$, the joint centrality appears to limit to one arc of a sinusoidal curve. We prove this result in Sec.~\ref{sec:asym}.
The most striking results occur in the bottom row of the third column. Based on conditional node  centrality, we see that MIT becomes the top-ranked university in the 1950s and then remains so in our data set.
Stanford and UC Berkeley develop gradually larger conditional centralities over the 64 years in the data set, whereas those of Princeton and Harvard decrease gradually over this period. When considering all universities, these five universities have conditional node centralities in the top-10 values throughout all years of the data set. This is consistent with our results in \cite{taylor2017eigenvector,taylor2019tunable}. 

We now examine directed interlayer coupling, and we take $\tilde{\bm{A}}$ to correspond to a directed chain with layer teleportation. See Eq.~\eqref{eq:tele} for the specific formula and the bottom row of Fig.~\ref{fig:toy1} for an associated visualization. In each panel of Fig.~\ref{fig:dir1}, we plot the joint centralities and conditional node centralities. The columns give results for interlayer coupling strengths of $\omega\in\{0.1,1,10,100\}$, and the three panels indicate different choices for the layer-teleportation probabilities: (A) $\gamma=0.0001$; (B) $\gamma=0.001$; and (C) $\gamma=0.01$. The dotted black curves in the rightmost column indicate large-$\omega$ asymptotic approximations that we will present in Section \ref{sec:asym}.

To understand the main effect of directed interlayer coupling, we first compare the joint centralities in Fig.~\ref{fig:dir1} to those in Fig.~\ref{fig:undir}. To help our discussion, we focus on the rightmost column of both figures. Comparing Fig.~\ref{fig:dir1} to Fig.~\ref{fig:undir}, we observe that the joint-centrality trajectories tend to decay with time for directed interlayer coupling, whereas they are peaked and attain their largest values near $t= 1978$ for undirected interlayer coupling. Therefore, directed interlayer coupling tends to ``boost'' the joint centralities of earlier time layers in comparison to undirected coupling. Comparing panels (A)--(C) of Fig.~\ref{fig:dir1} (and again focusing on the rightmost column), we observe that the rate of decay is fastest for $\gamma=0.0001$ (panel (A)) and slowest for $\gamma=0.01$ (panel (C)).  
The conditional centralities are also affected by directed interlayer coupling. Consider $\omega=10$ in Fig.~\ref{fig:undir}, and observe that the conditional centrality of Princeton decreases monotonically in time. By contrast, observe in Fig.~\ref{fig:dir1}(A,B) for $\omega=10$ that the conditional centrality of Princeton now decreases between $t=1946$ and $t \approx 1988$, but then it increases. 

For our last experiment, we examine how reversing the direction of interlayer edges changes the results of our supracentrality calculations. Specifically, we repeat the previous experiment with directed interlayer edges, except that now we set $\tilde{\bm{A}}$ to be the transpose of the matrix that we defined by Eq.~\eqref{eq:tele}.  One motivation is that for some applications, the most recent time layers are
more important than the earliest time layers. One can incorporate this idea into our supracentrality framework by reversing the direction of interlayer edges.
In Fig.~\ref{fig:dir2}, we plot the same quantities as in Fig.~\ref{fig:dir1}, except that now we take the directed interlayer edges to have the opposite direction (so we have reversed the arrow of time). Observe that the joint centralities now tend to increase with time, as opposed to Fig.~\ref{fig:dir1}, where they tended to decrease with time. These trends are most evident in the rightmost columns. We also observe differences for the conditional centralities. For example, focusing on $\omega=10$ in the third column of Fig.~\ref{fig:dir2}, we see that Princeton never has the largest conditional centrality. By contrast, for $\omega=10$ in Fig.~\ref{fig:undir} and Fig.~\ref{fig:dir1}(A,B), Princeton has the largest conditional centrality for the earliest time steps (specifically, for $t\in\{1946,\dots,1954\}$).

Understanding how the weights, topologies, and directions of interlayer coupling affect supracentralities is essential to successfully deploying supracentrality analysis to reveal meaningful insights.  
The above experiments highlight that one can tune the weights and topology of interlayer coupling to emphasize either earlier or later time layers. Specifically, one can adjust the parameters $\omega$ and $\gamma$, as well as the direction of interlayer edges, to cater a study to particular data sets and particular research questions. In our investigation in this section, we considered both the case in which $\tilde{\bm{A}}$ is given by Eq.~\eqref{eq:tele} and that in which it is given by the transpose of the matrix that we determine from Eq.~\eqref{eq:tele}. It is worth considering how these different choices of interlayer edge directions are represented in the supracentrality matrix $\mathbb{C}(\omega)$ and the consequences of these choices. Specifically, each layer's PageRank matrix ${\bf C}^{(t)}$ is defined in Eq.~\eqref{eq:google} using the transpose of the layer's adjacency matrix ${\bf A}^{(t)}$, yet when coupling the centrality matrices, we do not take the transpose of $\tilde{\bm{A}}$ when defining $\mathbb{C}(\omega)$ in Eq.~\eqref{eq:supracentrality}. Accordingly, one may worry that the matrix $\mathbb{C}(\omega)$ effectively acts in the forward direction for the intralayer edges, but in the opposite direction for the interlayer edges. However, this does not lead to any inherent contradiction, as the meanings of the directions in these two types of edges are fundamentally different: the direction of intralayer edges dictates the flow of random walkers, whereas the direction of interlayer edges couples the centralities of the different layers. In other applications, it may be necessary to encode the directions of the interlayer and intralayer edges in the same way, but there is no reason why one cannot encode directions of interlayer and intralayer edges in  different ways in a supracentrality formalism. As we have demonstrated by considering both $\tilde{\bm{A}}$ and its transpose --- and thus by treating the effect of the interlayer edges in opposite ways in these two calculations --- both uses are meaningful. They also probe different aspects of   temporal data. 
\section{Asymptotic Behavior for Small and Large Interlayer-Coupling Strength $\omega$}\label{sec:asym}

In this section, we summarize  the asymptotic results from \cite{taylor2019tunable} that reveal the behavior of supracentralities in the limit of small and large $\omega$. In our present discussion, we focus on dominant right eigenvectors.

To motivate our asymptotic analysis,
consider the top-right subpanels in each panel of Figs.~\ref{fig:undir}, \ref{fig:dir1}, and \ref{fig:dir2}. In each of these subpanels, we plot (in dotted black curves) the results of an asymptotic analysis of the dominant right eigenvector $\tilde{{\bf v}}^{(1)}$ of $\tilde{\bm{A}}$ for the joint centrality of MIT in the limit of large $\omega$. We observe excellent agreement with our numerical calculations.
Therefore, for sufficiently large $\omega$, one can understand the effects of both undirected and directed interlayer couplings (as encoded in an interlayer-adjacency matrix $\tilde{\bm{A}}$) by examining the dominant right eigenvector of $\tilde{\bm{A}}$. For large values of $\omega$, this eigenvector captures the limit of the joint centralities as a function with a peak
 for undirected coupled (see Fig.~\ref{fig:undir}), decay in time for directed coupling (see Fig.~\ref{fig:dir1}), and growth in time for directed coupling when reversing the arrow of time
 (see Fig.~\ref{fig:dir2}).  



\subsection{Layer Decoupling in the Limit of Small 
 $\omega$}

%

We begin with some notation. Let $\tilde{\mu}_1$ be the dominant eigenvalue (which we assume to be simple) of $\tilde{\bm{A}}$, and let $\tilde{\bm{u}}^{(1)}$ and $\tilde{\bm{v}}^{(1)}$ denote its corresponding left and right  eigenvectors. 
 Given a set $\{\mathbf{C}^{(t)}\}$ of centrality matrices, we let $\mu^{(t)}_1$ be the dominant eigenvalue (which we also assume to be simple) of   $\mathbf{C}^{(t)}$; and ${\bf u}^{(1,t)}$ and ${\bf v}^{(1,t)}$ are the corresponding left and right eigenvectors. Let $\{\mu^{(t)}_1\}$ denote the set of spectral radii, where $\lambda_{\rm{max}}(0) = \max_t \mu^{(t)}_1$ is the maximum eigenvalue over all layers. (Recall that $\lambda_{\rm{max}}(\omega)$ is the dominant eigenvalue of the supracentrality matrix $\mathbb{C}(\omega)$.) 
Let $\mathcal{P} = \{t : \mu_1^{(t)} = \lambda_{\rm{max}}(0) \}$ denote the set of layers whose centrality matrices achieve the maximum. 
When the layers' centrality matrices $\{\mathbf{C}^{(t)}\}$ are PageRank matrices given by Eq.~\eqref{eq:google}, it follows that $\mu_1^{(t)} =1$ for all $t$ (i.e., $\mathcal{P}=\{1,\dots,T\}$), the corresponding left eigenvector is ${\bf u}^{(1,t)} =[1,\dots,1]^T/N$, and $ {\bf v}^{(1,t)}$ is the PageRank vector for  layer $t$.
Furthermore, for each $t$, we define the length-$NT$ ``block'' vector ${\mathbbm{v}}^{(1,t)} = {{\bf e}}^{(t)} \otimes {\bf v}^{(1,t)}$, which consists of zeros except for block $t$, which equals ${\bf v}^{(1,t)}$.  (The vector ${{\bf e}}^{(t)}$ is a length-$T$ unit vector that consists of zeros except for entry $t$, which is $1$.)

We now present a theorem from \cite{taylor2019tunable}, although we restrict our attention to the part that describes the right dominant eigenvector.


\begin{theorem}[Weak-Coupling Limit of Dominant Right Eigenvectors \cite{taylor2019tunable}] \label{thm:uncoupled2}
Let 
$\mathbbm{v} (\omega)$ 
be the dominant 
right eigenvector of a supracentrality matrix that is normalized using the $1$-norm and satisfies the assumptions of Thm.~\ref{thm:unique}. Additionally, let $\mathcal{P} = \{t : \mu_1^{(t)} = \lambda_{\rm{max}}(0) \}$ denote the set of indices associated with the eigenvalues of $\mathbf{C}^{(t)}$ that equal the largest eigenvalue $\lambda_{\rm{max}}(0)$ of $\mathbb{C}(0)$. We assume that each layer's dominant eigenvalue $\mu_1^{(t)}$ is simple. It then follows that the
 $\omega \to 0^+$ limit of 
 $\mathbbm{v} (\omega)$   
satisfies
\begin{align}\label{eq:lim_uv}
	\mathbbm{v} (\omega)  \to \sum_{t\in\mathcal{P}}   \alpha_{t} \mathbbm{v}^{(1,t)}\,,  \quad
\end{align}
where the vector $\bm{\alpha} = [\alpha_{1},\dots,\alpha_{T}]^T$ has nonnegative entries 
and is the unique solution to the dominant eigenvalue equation
\begin{align}\label{eq:baah}
	\bm{X} \bm{ \alpha} &= \lambda_{1}  \bm{ \alpha}\,.
\end{align}
The eigenvalue $\lambda_{1}$ 
needs to be determined, and the entries of $\bm{X}$ are
\begin{align}\label{eq:Xtt}
	{X}_{tt'} &= \tilde{ {A}}_{t,t'}  \frac{ \langle {\bf u}^{(1,t)} , {\bf v}^{(1,t')} \rangle}{\langle{\bf u}^{(1,t)},   {\bf v}^{(1,t)} \rangle }   \chi(t)\chi(t') \,,
\end{align}
where $\chi(t)=\sum_{t'\in\mathcal{P}} \delta_{tt'}$ is an indicator function: $\chi(t)=1$ if $t\in\mathcal{P}$ and $\chi(t)=0$ otherwise.
The vector $\bm{\alpha} $ must also be normalized to ensure that the right-hand side of Eq.~\eqref{eq:lim_uv} is normalized (by setting $\|\bm{\alpha} \|_p=1$ for normalization with a $p$-norm). 
 \end{theorem}

\subsection{Layer Aggregation in the Limit of Large $\omega$}\label{sec:strong}
%


To study the $\omega\to\infty$ limit, it is convenient to divide Eq.~\eqref{eq:eig_eq} by $\omega$ and define $\epsilon=1/\omega$ to obtain
\begin{equation}\label{eq:newC}
	\tilde{\mathbb{C}}(\epsilon)  = \epsilon \mathbb{C}(\epsilon^{-1}) = \epsilon\hat{\mathbb{C}} +   \hat{\mathbb{A}}\,, 
\end{equation}
which has right eigenvectors  $\tilde{\mathbbm{v}}(\epsilon)$ that are identical to those of $\mathbb{C}(\omega) $ 
(specifically,   $\tilde{\mathbbm{v}}(\epsilon)={\mathbbm{v}}(\epsilon^{-1})$). Its eigenvalues $\{ \tilde{\lambda}_i\}$ are scaled by $\epsilon$, so $ \tilde{\lambda}_i(\epsilon) = \epsilon\lambda_i(\epsilon^{-1})$. 

Before presenting  results from \cite{taylor2019tunable}, we define a few additional concepts. Let $\tilde{\mathbbm{v}}^{(1,j)}=\tilde{{\bf e}}^{(j)} \otimes \tilde{\bm{ v}}^{(1)}$ denote a block vector that consists of $0$ entries, except for block $j$, which equals the dominant right eigenvector $\tilde{\bm{ v}}^{(1)}$ of $\tilde{\bm{A}}$.  
(The vector $\tilde{{\bf e}}^{(j)}$ is a length-$N$ unit vector that consists of $0$ entries except for entry $j$, which is $1$.) We also define the stride permutation matrix 
\begin{equation}\label{eq:stride}
	[\mathbb{P}]_{kl} = \left\{ \begin{array}{rl}
1\,,& \,\,l=\lceil k/N\rceil+T\,[(k-1)\bmod N]\,, \\
 0\,,& \,\,\textrm{otherwise} \, , 
\end{array} \right.
\end{equation}
where the ceiling function $\lceil \theta \rceil$ denotes the smallest integer that is at least $\theta$, and `$\bmod$' denotes the modulus function (i.e., $a \bmod b= a - b \lceil a/b -1\rceil$).

\begin{theorem}[Strong-Coupling Limit of Dominant Eigenvectors \cite{taylor2019tunable}] \label{thm:singular_limit}
Let $\tilde{\bm{A}}$, $\tilde{\mu}_1$, $\tilde{\bm{u}}^{(1)}$, and $\tilde{\bm{v}}^{(1)}$ be defined as above, with the same assumptions as in Theorem~\ref{thm:unique}.
It then follows that the dominant eigenvalue $\tilde{\lambda}_{\rm{max}}(\epsilon)$ and the associated eigenvector $\tilde{\mathbbm{v}}(\epsilon)$ of $\tilde{\mathbb{C}}(\epsilon)$ converge as $\epsilon \to 0^+$ to the following expressions: 
\begin{align}
	\tilde{\lambda}_{\rm{max}}(\epsilon)     &\to\tilde{ \mu}_1   \,,~~ \nonumber \\ 
	\tilde{\mathbbm{v}}(\epsilon) &\to \sum_j \tilde{\alpha}_{j} \mathbb{P}\tilde{\mathbbm{v}}^{(1,j)} \,, 	\label{eq:conver2}
\end{align}
where the constants   $\{\tilde{\alpha}_i\}$ solve the  dominant eigenvalue equation 
\begin{equation}\label{eq:alpha_beta}
	\tilde{\mathbf{X}}  \tilde{\bm{\alpha}} =  \tilde{\mu}_1 \tilde{\bm{\alpha}}\,,  
\end{equation}
with 
\begin{align}
	\tilde{X}_{ij} 	
		&=   \sum_{t} {C}^{(t)}_{ij} \frac{\tilde{u}_t^{(1)} \tilde{v}_t^{(1)}}{\langle \tilde{\bm{u}}^{(1)}, \tilde{\bm{v}}^{(1)} \rangle} \,. \label{eq:M1star_0_q} 
\end{align} 
Note that we normalize the vector $\tilde{\bm{\alpha}} $ 
to ensure that the right-hand side of Eq.~\eqref{eq:conver2} is normalized.
\end{theorem}


%



Equation~\eqref{eq:M1star_0_q} indicates that the strong-coupling limit effectively aggregates the centrality matrices $\{ {\bf C}^{(t)} \}$ across time via a weighted average, with weights that depend on the dominant left and right eigenvectors of $\tilde{\bm{A}}$. When $\tilde{\bm{A}}$ encodes an undirected chain from Eq.~\eqref{eq:undir} (see the top row of Fig.~\ref{fig:toy1}), it follows that \cite{taylor2017eigenvector}
\begin{equation}\label{eq:sin}
	\tilde{\mathbf{X}} =   \sum_{t} \mathbf{C}^{(t)}  \frac{\sin^2\left( \frac{\pi t}{T+1}\right)}{\sum_{t=1}^T \sin^2\left( \frac{\pi t}{(T+1)}\right)}\,.
\end{equation}	
The dotted black curve in the top-right subpanel of Fig.~\ref{fig:undir} shows a scaled version of $ \tilde{\bm{v}}^{(1)} $, which is defined by the normalized sinusoidal weightings in Eq.~\eqref{eq:sin}. 
The dotted black curves in the top-right subpanels of each panel of Figs.~\ref{fig:dir1} and \ref{fig:dir2} also show $ \tilde{\bm{v}}^{(1)} $ (specifically, when $\tilde{\bm{A}}$ is given by Eq.~\eqref{eq:tele} or by the transpose of the matrix that we obtain from Eq.~\eqref{eq:tele}, respectively), which we scale  to normalize the joint centralities.

\section{Discussion}

We   presented a supracentrality framework to study how the importances of nodes in a temporal network change over time. Our approach involves representing a temporal sequence of networks as time layers in a multiplex network and using the strength and topology of coupling between time layers to tune centrality trajectories. A key feature of our approach is that it simultaneously yields the centralities of all nodes at all times by computing the dominant right eigenvector of a supracentrality matrix.

Inspired by ideas from probability theory, we examined three types of eigenvector-based supracentralities:
\begin{enumerate}
\item[(i)] the joint centrality for a node-layer pair $(i,t)$; this captures the combined importance of node $i$ and time layer $t$;
\item[(ii)] the marginal centrality of node $i$ or time $t$; these captures separate importances of a node or a time layer; and
\item[(iii)] the conditional centrality of a node $i$ at time $t$; this captures the importance of a node relative only to other nodes at that particular time.
\end{enumerate}

Because our approach involves analyzing the dominant eigenvector of a centrality matrix, it generalizes 
eigenvector-based centralities,  such as PageRank, hub and authority centralities, and (vanilla) eigenvector centrality.  Naturally, it is desirable to extend supracentralities to analyze networks that are both temporal and multiplex \cite{kivela2014}. Another important generalization of centrality analysis is the study of continuous-time temporal networks and streaming network data \cite{Grindrod_Higham_2014,walid2018}, and it would be insightful to extend   supracentralities to such situations.

 
 \section*{Acknowledgements}
We thank Petter Holme and Jari Saram{\"a}ki for the invitation to write this chapter. We thank Deryl DeFord, Tina Eliassi-Rad, Des Higham, Christine Klymko, Marianne McKenzie, Scott Pauls, and Michael Schaub for fruitful conversations. 
DT was supported by the Simons Foundation under Award \#578333.
PJM was supported by the James S. McDonnell Foundation 21st Century Science Initiative --- Complex Systems Scholar Award \#220020315.

\bibliographystyle{spphys}
\bibliography{supracentrality2}

\end{document}